\documentstyle[12pt]{article}
\newcommand{\subeqn}[1]{\newcounter{eqn}\addtocounter{equation}{1}
\setcounter{eqn}{\theequation}\setcounter{equation}{0}
\renewcommand{\theequation}{\theeqn.\arabic{equation}}
\begin{eqnarray}#1\end{eqnarray}\setcounter{equation}{\arabic{eqn}}
\renewcommand{\theequation}{\arabic{equation}}}
\begin{document}


\begin{titlepage}
\title{\bf The Equivalence Theorem\\and\\infrared divergences}
\author{Tibor Torma\protect\thanks{e--mail: kakukk@phast.umass.edu}}
\date{\today}
\maketitle
\begin{center}{hep-ph/9510217}\end{center}
\vspace{1in}
\begin{abstract}
We look at the Equivalence Theorem as a statement about the absence
of polynomial infrared divergences when $m_W \rightarrow 0$. We prove their
absence in a truncated toy model and conjecture that, if they exist at all,
they are due to couplings between light particles.
\end{abstract}
\renewcommand{\thepage}{\mbox{}}
\end{titlepage}

\pagenumbering{arabic}
\renewcommand{\thesection}{\Roman{section}}
\section{\bf Introduction}

The Equivalence Theorem~\cite{eqth}, from a practical point of view, is a
tool to simplify calculations in spontanously broken gauge theories. It states
that, in the limit $\frac{\sqrt{s}}{m_W}\rightarrow\infty$, amplitudes with
external longitudinal $W_L$'s and $Z_L$'s can be replaced by the corresponding
(unphysical) Goldstone boson amplitudes. With this replacement one gains a
simplification in Lorentz indeces and calculations are simplified by the lack
of gauge cancellations.

The early incomplete proofs of the Equivalence Theorem within the
standard model~\cite{eqth} were improved using power counting
arguments~\cite{GrKet,HVelt}; see also~\cite{He}. One distinguishes two
regimes, the light--Higgs regime with $\sqrt{s} >> m_H, m_W, m_f$ and a
heavy--Higgs regime with $m_H\sim\sqrt{s} >> m_W, m_f$;
in both cases excluding situations with exceptional momenta
($(p_i-p_j)^2<<E^2$) which could upset the power counting. The power
counting proof of the Equivalence Theorem as given in~\cite{GrKet}
works to all orders; it also shows that in the heavy--Higgs regime
all Feynman diagrams with lines in loops other than Goldstone scalar
are subleading, allowing for a consistent truncation of Feynman
amplitudes to pure Goldstone dynamics.

In this paper we discuss a possible flaw in the power counting proof
which can occur if there are power-like infrared divergences. We
argue that these do not occur in the minimal Higgs sector of the
Standard Model, and use a toy model to argue that
they are probably not present in its multi-Higgs extensions either.

C.~Grosse-Knetter's argument~\cite{GrKet} involves counting powers of
$m_H$ and $E=\sqrt{s}$ simultaniously. One starts with the well-known
statement that
\(<phys\mid TF_1\dots F_n\mid phys>\;=0\)
for the $R_\xi$-gauge fixing terms
\(F_i=A^\mu_i-ig\xi\cdot\partial^\mu\Phi_i\rightarrow 0\), connecting
matrix elements of the Goldstone bosons
$\phi_i$ to vector bosons with (unphysical) polarization
$\propto\frac{p^\mu}{m_V}$. Then one is to
prove that the difference between the polarization vectors of these $V$'s and
 the longitudinal $V_L$ vector bosons,
\(v^\mu=\epsilon^\mu-\frac{p^\mu}{m_V}\equiv\frac{m_V}{E_V+p_V}\cdot(1\mid\bf
0)^\mu\) gives only subleading contributions in the appropriate limit.
A generic Feynman amplitude is, explicitely displaying the $E$ and $m_H$
factors, a sum of terms, each in the form
\begin{equation}
{\cal M} =c\cdot E^{\frac{E_f}{2}-E_v}\cdot
m_H^{2V_\Phi}\cdot I_F,
\end{equation}
where the constant $c$ may depend on 
the $m_V$'s, $V_\Phi$ is the added number of $\Phi^3$ and $\Phi^4$ Higgs
self-interactions, $E_f$ is the number of external fermions and $E_v$ is the
number of $v^\mu$'s; the remaining part of the Feynman amplitude is in the
general form
\begin{equation}\label{IF}
I_F=\int d^{4L}k \cdot \frac{p\dots p}{(q_1^2-m_1^2)
\dots(q_I^2-m_I^2)},
\end{equation}
where we collectively denoted by $p$ and $q$ the occurring linear
combinations of external and loop momenta; $m$ denotes both the heavy
and the light massses.

Eqn.~\ref{IF} is a general formula for all processes and should
include all internal lines present in the theory. As explained below,
an assumption on its behavior in general at large scales can be used
to prove the Equivalence Theorem. For this purpose, the relevant
graphs are those which include at least one external longitudinal
vector boson.

If we now suppose that $I_F$ is determined by the scale $E$ and $m_H$, we have
\begin{equation}
I_F =m_H^D\cdot f_0(\frac{E}{m_H})+
 m_H^{D-1}\cdot m_W\cdot f_1(\frac{E}{m_H})+\dots\label{eq:feyn}
\end{equation}
with dimensionless $f_j$ (possibly containing logarithms of $m_W$ and of the
renormalization scale but no powers of $\frac{1}{m_W})$ easy combinatorics 
shows that the total power of $E$ and $m_H$, counted simultaniously, in ${\cal
M}$ is at  most
\begin{equation}
N=(2L+2)-(V_d+2V_0+V_f+E_v)
\end{equation}
where $L$ is the number of loops, $V_d$ and $V_f$ are the numbers of
derivative and fermion couplings respectively; $V_0=V-V_\Phi-V_d-V_f$.
This formula shows that the leading graphs are, at each loop level, those
with $E_v=0$ (i.e. proves the Equivalence Theorem) and also shows that no
vector or fermion lines are involved in loops in leading graphs. The former
statement can also be proven true for the light-Higgs regime using a similar 
argument.

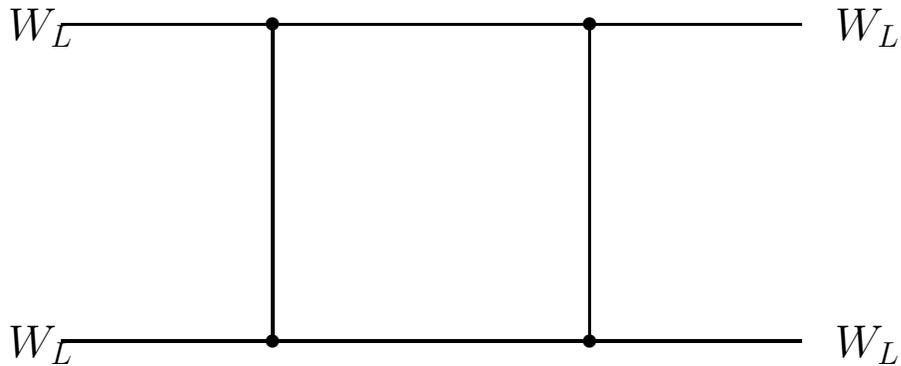
\begin{figure}[htb]
\setlength{\unitlength}{.4pt}
\begin{picture}(972,500)
\thicklines
\put(150,400){\line(1,0){700}}
\put(150,100){\line(1,0){700}}
\put(350,100){\circle*{12}}\put(350,400){\circle*{12}}
\put(650,100){\circle*{12}}\put(650,400){\circle*{12}}
\put(350,100){\line(0,1){300}}
\put(650,100){\line(0,1){300}}
\thinlines
\put(100,385){\shortstack{\Large $W_L$}}
\put(882,385){\shortstack{\Large $W_L$}}
\put(100,85){\shortstack{\Large $W_L$}}
\put(882,85){\shortstack{\Large $W_L$}}
\end{picture}
\caption{One of the leading graphs that might
pick up IR divergences at 1-loop level.}\label{fig0}
\end{figure}

A possible flaw in this argument can occur if negative powers of the
light mass $m_V$ enter the expansion~(\ref{eq:feyn}). That expression
can be viewed as an infrared statement: up to possible logarithmic
factors involving the renormalization scale, we can fix the unit of
dimension at~$E$ (or at~$m_h\propto E$), to get a theory with
$m_V\rightarrow 0$, wherein $m_V$ acts as an infrared regulator.
Then, the statement in~(\ref{eq:feyn}) has been transformed into the
statement that {\it the leading part of a graph, $f_0$, does not pick
up polynomially divergent factors when $m_V\rightarrow 0$}. In many
calculations of IR divergent quantities that use a small regulator
mass (for examples, see~\cite{Zwanziger,Weinb} among many others) one
finds that the divergences are only logarithmic in the regulator mass
and this is the basis for the expansion in~(\ref{eq:feyn}). Any
infrared divergences of the polynomial type in $\frac{1}{m_V}$ would
introduce additional factors of $\frac{m_H}{m_V}$ and/or
$\frac{E}{m_V}$ and such terms would invalidate the above proof which
is based on their absence. To see if any such factor actually upsets
the Equivalence Theorem one needs to analyse by how many powers of
$m_V$ the graph in question is suppressed in~(\ref{eq:feyn}). For
exemaple, $V_LV_L\to V_LV_L$ at one loop has
$\left(\frac{m_W}{E\mbox{ or }m_W}\right)^4$, i.e. $N=-4$ as leading
behavior. Any additional factor of $\frac{1}{m_V}$ in, say, the box
diagram of Fig.~\ref{fig0}, which itself is one of the leading graphs,
would introduce a factor that breaks the Equivalence Theorem.

 A related problem could occur in a theory with several Higgses when
some of them are heavy and at least one is light. Powers of
$\frac{1}{m_h}$ for the light Higgs could upset the power counting.

What the argument in~\cite{GrKet} does is, in this light, that it
shifts the question of 'breaking' the Equivalence Theorem to the
question of the presence of severe (i.e. polynomial) IR divergences.

This point of view is in accordance with the view that the Equivalence Theorem
expresses the fact that Goldstone d.o.f's are turned into longitudinal vector
bosons by a spontanously broken gauge transformation. With fixed $E$ and
$m_V \rightarrow 0$ we approach the point of phase transition where the
t'Hooft gauge condition \(\partial_\mu V^\mu=m_W\cdot\phi\) turns from a way 
to express $\phi$ with $V_L$ into the physicality condition on $V_L$.
The~$m\rightarrow 0$ limit of a vector boson is a notoriously tricky problem.
Whether the transition is smooth enough not to break the relations between
amplitudes or not, will show up in the presence or lack of 'bad' IR 
divergences.

In this paper we argue that such divergences do not occur in the pure
Higgs sector, at least, for non-exceptional momenta. We try to relate
the IR divergence found in scalar box graphs (see Eqn.~(\ref{eq:sing}) below
to the Equivalence Theorem and find no such divergence when the external
particles are massive. In the copnclusion we point out that a similar
treatment is necessary for massless external particles which requires an
analysis more similar to the case of the IR divergencesd in QCD. Experience
with graphs involving massless particles of different spins shows that the
infrared divergences of a diagram are generally determined by the
denominators in the integrals of Eqn.~(\ref{IF}) and the presence of
spin affects only the numerators in that expression. The result is
that the general structure of the IR divergences is independent of the
spin (see e.g.~\cite{Zwanziger,Weinb}) and the reason is the following.
IR divergences come from specific points in the loop integration
region where 4-momenta $k^\mu$ of some massless propagators vanish.
One can then expand the numerators around those points in $k$. We see
that in terms of higher order in $k$, IR divergences cancel and the
remaining divergent part involves only a constant, i.e. a structure
that is similar to the spin-0 case. This argument tells us that the
absence of IR divergences in the scalar sector essentially proves
the absence of divergences in the full model. One might hope that for
some theories one could get special behavior from the numerators for
special reasons, such as gauge invariance, that might remove an IR
divergence that would otherwise be present. However, the scalar theory
then would only be worse, so that by exploring the scalar theory one
should be able to find the strongest divergences.

The situation here is analogous to QED in that interactions of massive
particles through massless photons (the latter now correspond to the
W and Z) introduce only logarithmic IR divergences. To complete the
proof of the Equivalence Theorem to all orders one should also prove
that in a situation more reminescent of QCD, where massless gluons
(now corresponding to W and Z) introduce IR divergences that are much
harder to handle, polynomial IR divergences are also absent -- a
question we do not address here.

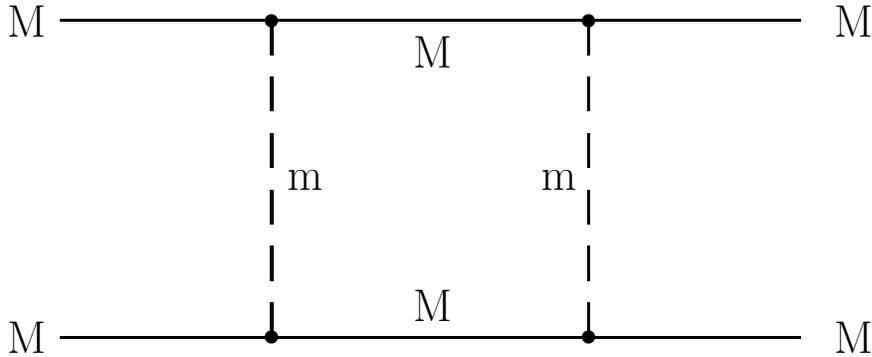
\begin{figure}[htb]
\setlength{\unitlength}{.4pt}
\begin{picture}(972,500)
\thicklines
\put(150,400){\line(1,0){700}}
\put(150,100){\line(1,0){700}}
\put(350,100){\circle*{12}}\put(350,400){\circle*{12}}
\put(650,100){\circle*{12}}\put(650,400){\circle*{12}}
\multiput(350,100)(0,53.5){6}{\line(0,1){32.5}}
\multiput(650,100)(0,53.5){6}{\line(0,1){32.5}}
\thinlines
\put(484,355){\shortstack{\Large M}}
\put(484,115){\shortstack{\Large M}}
\put(100,385){\shortstack{\Large M}}
\put(882,385){\shortstack{\Large M}}
\put(100,85){\shortstack{\Large M}}
\put(882,85){\shortstack{\Large M}}
\put(365,240){\shortstack{\Large m}}
\put(605,240){\shortstack{\Large m}}
\end{picture}
\caption{One-loop graphs with powerlike $\frac{1}{m}$ behavior.}\label{f1}
\end{figure}

The form of IR divergences of Feynman graphs has been calculated in QED
many times (see, e.g.~Zwanziger~\cite{Zwanziger}), and is usually found to be
logarithmic in the various IR regulators to any finite order. It has been
calculated~\cite{Weinb} for gravitons with similar results. However, it is
well-known that in the nonrelativistic limit scalar box graphs, such as that
in Fig.~\ref{f1}. pick up $O\left(\frac{1}{m^2}\right)$ terms for forward
scattering~\cite{Gross}:
\begin{equation}\label{eq:sing}
i{\cal M}{\sim}\frac{1}{m^2M(p+\frac{i}{2}m)}
\end{equation}
clearly showing the type of behavior expected to break the Equivalence 
Theorem. It is worth to note, however, that these calculations use either a
small  off-shellness
parameter~\cite{Zwanziger} \(\varepsilon_i = p_e^2 - m_e^2 \) or an explicit
cutoff~\cite{Weinb} \(\mid p_\gamma^0\mid<\lambda\) as an IR regulator, so
they are not directly relevant to our case where a small regulator mass $m$
should be used.

It is an interesting co\"\i ncidence that the structure
in~Fig.~\ref{f1}. is not present in the Standard Model. The only heavy
mass there is $M\rightarrow m_H$ and there are no $HHW$ or $HHZ$ couplings
(the absence of the latter is a consequence of separate $C$ and 
$P$-conservation
in the purely bosonic part of the SM, where $J^{PC}(Z)=1^{--}$). All other box
graphs are less IR divergent because less interal lines can be put on shell
in the same time.

The obvious way to look for a similar divergence is in the two--doublet
Higgs model which posesses one heavy Higgs $M\rightarrow m_H$ and at least
one light $0^{++}$ Higgs $m\rightarrow m_h\leq m_Z$. Even though such an IR
divergence is not a genuine $\frac{1}{m_W}$, because of the necessary
lightness of $m_h$, it affects the amplitudes in the same way.

In Sect.~\ref{sWeinb} we use Weinberg's method to analyse~\cite{Weinb} the
'truncated MSSM model' to all orders, dropping all particles from there 
except  the heavy
and the light $0^{++}$ Higgses, using $m\rightarrow  m_h$ as an infrared
regulator. This model conserves the number of heavy $H$'s, which so act as
'charged' particles while the light $h$'s act as uncharged scalar photons.
The lack of the coupling between light particles allows to extract the IR
divergences in complete analogy to the QED case and we find only
logarithmic divergences. The $S$-matrix elements pick up factors
\begin{equation}
S_{\beta\alpha} = S_{\beta\alpha}^{(0)}\cdot
\exp{\left\{\frac{g^2}{4(2\pi)^2}\cdot({\cal G}+i{\cal F})\cdot
\log{\frac{\Lambda}{m}}\right\}}\label{e4}
\end{equation}
where $2\pi g=Gm_Z$ is an MSSM coupling constant, $m\ll\Lambda\ll E$ is an 
energy
when we separate soft particles from hard ones. ${\cal F}$ has contributions
from pairs of incoming and from pairs of outgoing $H$'s; ${\cal G}$ has
contributions from all pairs. As we will see in Sect.~\ref{sWeinb}, ${\cal G}$
is canceled on the cross section level by real soft $h$'s; the ${\cal F}$
does not contribute to the cross section but it shifts the Coulomb phase by
an $\sim \log{\frac{\Lambda}{m}}$ term, thus explaining why there is no
Coulomb phase contribution from a pair formed of one incoming and one
outgoing~$H$. In Sect.~\ref{sbox} we calculate explicitely the IR divergent
contributions in $HH\rightarrow HH$ to one-loop level and find complete
agreement with the above conclusions. It is of some interest how these IR
divergent terms are separated using our small-mass regularization: one needs
a careful and long procerdure which is illustrated by describing the details
for one particular diagram. The mechanism of the cancellation of IR 
divergences, all
logarithmic in $m$, between loop integrals and soft 'photons' attached to 
heavy external legs closely follows the corresponding mechanism in
$ee\rightarrow ee$ in QED, so it is straightforward to generalize the result of
Sect.~\ref{sWeinb} for any other process in this toy model, for example, for
$hH\rightarrow hH$ or $hh\rightarrow hh$, analogous to $e\gamma\rightarrow
e\gamma$ or $\gamma\gamma\rightarrow\gamma\gamma$.

The total absence of powerlike divergences shows that the $\frac{1}{m^2}$
divergence of the nonrelativistic forward amplitude is due to exceptional
momenta. This statement is in compliance with that the coefficient of our
$\log{\frac{\Lambda}{m}}$ divergence itself diverges when two particles are
collinear (see Eqn.~(\ref{eFandG}), \(\beta_{ij}\rightarrow 0\)).

In Sec.~\ref{sWeinb} we show that in a simplified model with two
scalars only no polynomial IR divergences occur. As a consistency
check we show that the IR structure found is identical with the one
necessary in order that soft Bremsstrahlung exactly cancels the IR
divergences in loops. We then illustrate this general argument in
Sect.~\ref{sbox} with a calculation of the relevant 1-loop amplitudes.
In the conclusion we point out that the presence of coupled light
particles (i.e. gauge boson self-coupling) introduces additional
difficulities which require a separate investigation.

\section{\bf The 'truncated MSSM' model}\label{sWeinb}

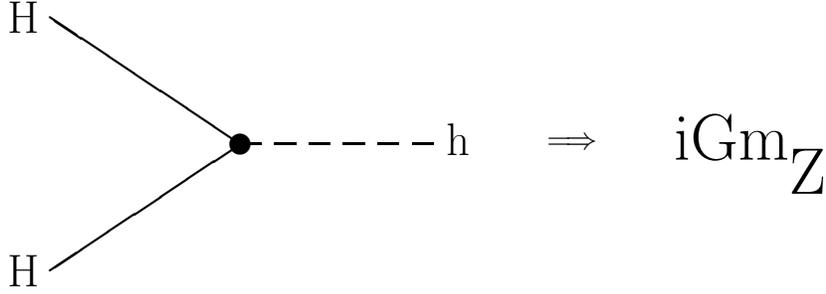
\begin{figure}[htb]
\setlength{\unitlength}{.4pt}
\begin{picture}(972,550)
\thicklines
\put(310,250){\circle*{20}}
\put(310,250){\line(-3,2){180}}
\put(310,250){\line(-3,-2){180}}
\multiput(310,250)(32.5,0){6}{\line(1,0){20}}
\thinlines
\put(90,355){\shortstack{\Large H}}
\put(90,115){\shortstack{\Large H}}
\put(505,240){\shortstack{\Large h}}
\put(600,245){\shortstack{$\Longrightarrow$}}
\put(720,235){\shortstack{$\mbox{\Huge iGm}_{\mbox{\Huge Z}}$}}
\end{picture}
\caption{The only remaining vertex in the truncated MSSM.}\label{f2}
\end{figure}

We use the $H$ and $h$ part of the MSSM as a toy model to illustrate why IR
divergences are at most powers of logarithms in the small masses. This model 
has only one coupling (see Fig.~\ref{f2}) with a dimensionless $G\sim
g_{weak}$; the $m_Z$ factor in the coupling should not be considered as
suppressing the IR divergence: {\it any} deviation in the expansion of $I_F$
from Eqn.~(\ref{eq:feyn}) upsets the power counting. In this model, with
$M\rightarrow m_H$ and E kept constant and $m\rightarrow m_h$ sent to zero,  
we calculate the IR divergences, closely following the argument in
Weinberg~\cite{Weinb}.

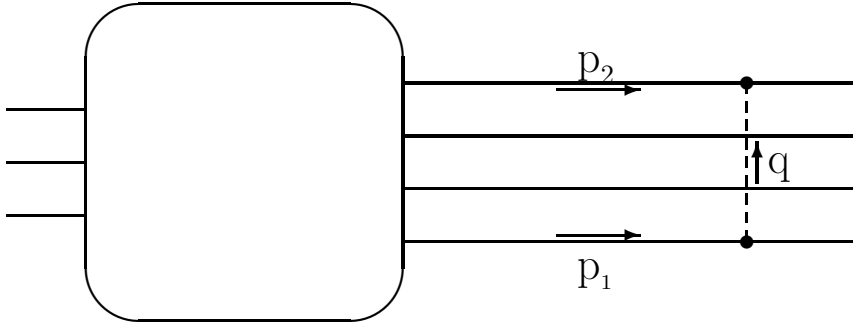
\begin{figure}[htb]
\setlength{\unitlength}{.4pt}
\begin{picture}(972,450)
\thicklines
\put(425,175){\line(1,0){425}}
\put(425,225){\line(1,0){425}}
\put(425,275){\line(1,0){425}}
\put(425,325){\line(1,0){425}}
\put(125,200){\line(-1,0){75}}
\put(125,250){\line(-1,0){75}}
\put(125,300){\line(-1,0){75}}
\put(750,175){\circle*{12}}
\put(750,325){\circle*{12}}
\multiput(750,175)(0,17.5){9}{\line(0,1){10}}
\put(275,250){\oval(300,300)}
\put(570,319){\vector(1,0){80}}
\put(570,181){\vector(1,0){80}}
\put(760,230){\vector(0,1){40}}
\thinlines
\put(770,240){\shortstack{\Large q}}
\put(590,140){\shortstack{${\mbox{\Large p}}_{\mbox{\scriptsize 1}}$}}
\put(590,335){\shortstack{${\mbox{\Large p}}_{\mbox{\footnotesize 2}}$}}
\end{picture}
\caption{Attaching a soft exchange to external lines.}\label{f3}
\end{figure}

All the IR divergences in an {\it amplitude} stem from a set of soft $h$ 
exchanges between external legs of Feynman graphs. Their
factorization~\cite{Weinb} is due to the fact that these corrections are
attached to all graphs in the same way. We have, for each attached soft
exchange (see Fig.~\ref{f3}.) \begin{equation}
S_{\beta\alpha}=S_{\beta\alpha}^{(0)}\cdot\left\{1+\int d^4q\cdot
A(q)\right\}
\end{equation}
and
\begin{eqnarray}
A(q)&=&\frac{i}{(2\pi)^4}\cdot\int\frac{d^4q}
{q^2-m^2+i\varepsilon}\cdot\\
&&\mbox{}\cdot\frac{i(i\cdot 2\pi g)}{(p_1+\eta_1q)^2-M^2+i\varepsilon}\cdot
\frac{i(i\cdot 2\pi g)}{(p_2-\eta_2q)^2-M^2+i\varepsilon}\nonumber
\end{eqnarray}
and $\eta_j=+1\ (\mbox{or}\ -1)$ when the $H$ is outgoing (incoming). 
Attaching all possible soft exchanges leads to a factorized
amplitude~\cite{Weinb} \begin{equation}
S_{\beta\alpha}=S_{\beta\alpha}^{(0)}\cdot\exp{\int d^4q\cdot A(q)}\ .
\end{equation}

The integration is over a \([-\Lambda,+\Lambda]\) range in $q^0$ with
\mbox{IR cut $\ll\Lambda\ll E$}; the resulting $\Lambda$ dependence is 
compensated by a $\Lambda$-dependence in the hard $h$ part
$S_{\beta\alpha}^{(0)}$. In addition, by the usual $\mid q^0\mid>2$ cutoff we
get, after some elementary integrations, with $m\equiv0$:
\begin{equation}
A_\lambda=\frac{g^2}{8}\cdot\sum_{j\neq l}\frac{\eta_j\eta_l}
{(p_j\cdot p_l)}\cdot\frac{1}{\beta_{jl}}\log{\frac{1+\beta_{jl}}{1-\beta_{jl}}}
\end{equation}
where $\beta_{jl}$ is the relative velocity of two $H$'s
\begin{equation}
\beta_{jl}=\sqrt{1-\frac{M^4}{(p_j\cdot p_l)^2}}\ .
\end{equation}

What we need is to use $m>0$ instead of $\lambda>0$; to calculate it we need  
a careful analysis of how complex singularities move around on the $q^0$ plane.
Closing the contour around them and integrating over angular variables leaves 
us with one-dimensional integrals. Separating their IR divergent (i.e. when
$m\rightarrow 0$) parts is a tedious calculation; we only quote the result in 
the form of Eqn.~(\ref{e4}) with

\begin{eqnarray}
{\cal G}&=&-2\pi^2\sum_{j\neq l}\frac{\eta_j\eta_l}
{(p_j\cdot p_l)\beta_{jl}}\cdot
\log{\frac{1+\beta_{jl}}{1-\beta_{jl}}}\nonumber\\
\mbox{and}&&\label{eFandG}\\
{\cal F}&=& \sum_{j\neq l\ \mbox{with}\ \eta_j=\eta_l}
\ \ \frac{(4\pi)^3}{(p_i\cdot
p_j)^2-M^4}\ .\nonumber
\end{eqnarray}

In a physical process, the
\(\exp{\left\{i\cdot\frac{g^2}{4}\cdot{\cal F}\cdot\log{\frac{\Lambda}{m}}
\right\}}\)
factor contributes to the Coulomb phase as we said in the introduction; the
{$\cal G$}
part goes into the cross section and gets compensated by real soft $h$'s.

One may always add any number of indetectably soft $h$'s to the initial and/or 
final states as long as their total energy is less then the energy resolution
of the measuring device. Although this has not much practical sense for
\mbox{$m_h\sim 10$'s of $GeV$'s}, we still see that a Bloch-Nordsieck-type
cancellation occurs. Attaching $N$ soft $h$'s to the external legs,
the amplitude picks up an IR divergent factor
\begin{equation}
S_{\beta\alpha}({\bf q}_1,\dots,{\bf q}_N)= S_{\beta\alpha}
\cdot\frac{1}{N!}\cdot\prod_{n=1}^{N}\sum_j\frac{i(i\cdot 2\pi g)}
{(p_j-\eta_j\cdot
q)^2-M^4}\ .
\end{equation}

These states should be added on the probability level as they represent
orthogonal states. Using the factorization property as in~\cite{Weinb} again,
we get for the transition rate
\begin{eqnarray}
\Gamma_{\beta\alpha}({\bf q}_1,\dots,{\bf q}_N)&=&
\mid S_{\beta\alpha}\mid^2\cdot\frac{1}{N!}\cdot\prod_{n=1}^{N}
\frac{g^2}{16\pi E_{q_n}}\cdot\\
&&\mbox{}\cdot\sum_{jl}\frac{\eta_j\eta_l}
{\left[(p_j\cdot q_n)+\eta_j\frac{m^2+i\varepsilon}{2}\right]
\left[(p_l\cdot q_n)+\eta_l\frac{m^2-i\varepsilon}{2}\right]}\ .\nonumber
\end{eqnarray}

Integrating this over the $h$'s' phase space with the restriction
\begin{equation}
\chi\left(\sum_{n=1}^{N}E_n\leq\Lambda\right)\equiv\frac{1}{\pi}\cdot
\int_{-\infty}^{+\infty}\frac{d\sigma}{\sigma}\cdot\sin{\sigma}\cdot
e^{i\frac{\sigma}{\Lambda}\cdot\sum_{n=1}^NE_n}
\end{equation}
we have, with the total energy carried away by the $h$' less than $\Lambda$:
\begin{eqnarray}
\lefteqn{\Gamma_{\beta\alpha}(\leq\Lambda)\ =\ 
\mid S_{\beta\alpha}\mid^2\cdot\frac{1}{\pi}\cdot
\int_{-\infty}^{+\infty}\frac{d\sigma}{\sigma}\cdot\sin{\sigma}\cdot}
\nonumber\\
&&\cdot\exp{}
\{\frac{g^2}{16}\cdot\int_0^\infty
\frac{dq\cdot q^2}{E_q}\cdot e^{i\frac{\sigma}{\Lambda}\cdot E_q}\cdot\\
&&\ \ \ \ \ \ \cdot\sum_{jl}\oint_{4\pi}
\frac{\eta_j\eta_l\cdot d^2\underline{n}}
{\left[(p_j\cdot q)+\eta_j\frac{m^2+i\varepsilon}{2}\right]\cdot
 \left[(p_l\cdot q)+\eta_l\frac{m^2-i\varepsilon}{2}\right]}        
\ \}\ .\nonumber
\end{eqnarray}

The separation of IR divergences requires a hard mathematical procedure that 
we do not present here; in essence, we write the integral
\begin{eqnarray}
\lefteqn{\int_0^\infty\frac{dq\cdot q^2}{E_q}\cdot e^{i\frac{\sigma}
{\Lambda}E_q} \Longrightarrow}\\
&&\Longrightarrow\int_\Lambda^\infty\frac{dq\cdot q^2}{E_q}\cdot
e^{i\frac{\sigma}{\Lambda}E_q}+\int_0^\Lambda\frac{dq\cdot q^2}{E_q}\cdot
\left(e^{i\frac{\sigma}{\Lambda}E_q}-1\right)+\int_0^\Lambda\frac{dq\cdot q^2}
{E_q} \nonumber
\end{eqnarray}
and prove that the first two terms do not contribute to the IR divergence.  
Our result is
\begin{equation}
\Gamma_{\beta\alpha}=\mid S_{\beta\alpha}\mid^2\cdot\exp{\left\{
-\frac{g^2}{4(2\pi)^2}\cdot{\cal G}\cdot\log{\frac{\Lambda}{m}}\right\}}
\end{equation}
with the same {$\cal G$} as in Eqn.~(\ref{eFandG}), proving the cancellation 
of all IR divergences in the transition probabilities.

Expanding any of our results in the coupling constant $G$ shows that, to all
orders, the worst divergence is only a power of \(\log{\frac{\Lambda}{m}}\).
We note that although summing up all orders we certainly get a powerlike 
behavior, \begin{equation}
S_{\beta\alpha}=S^{(0)}_{\beta\alpha}\cdot{\left(\frac{\Lambda}{m}\right)}^{
-\frac{g^2}{2(2\pi)^2}\cdot{\cal G}},
\end{equation}
we do not think this points to the breaking of the Equivalence Theorem; we 
are, after all, summing for a very particular set of diagrams.

\section{\bf A one-loop example}\label{sbox}

In order to illustrate the results of Sect.~\ref{sWeinb}, and also to see how 
the IR divergences of particular diagrams add up, we work out the infrared
divergences of $HH\rightarrow HH$ at one loop level and find complete agreement
with Eqns.~(\ref{e4},\ref{eFandG}). Fortunately, all graphs with IR divergences
are UV finite,
so we may ignore renormalization. It turns out that each individual divergent
graph has a \(\log{\frac{\Lambda}{m}}\) divergence and no cancellations occur.

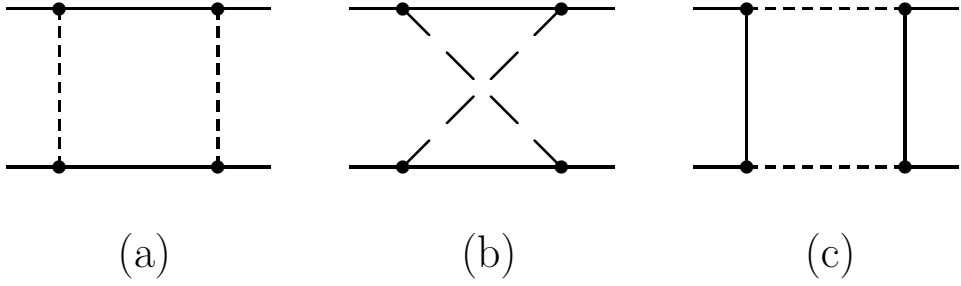
\begin{figure}[bht]
\setlength{\unitlength}{.4pt}
\begin{picture}(972,350)
\thicklines
\put(50,130){\line(1,0){250}}
\put(50,280){\line(1,0){250}}
\multiput(100,130)(0,17.5){9}{\line(0,1){10}}
\multiput(250,130)(0,17.5){9}{\line(0,1){10}}
\put(100,130){\circle*{12}}
\put(250,130){\circle*{12}}
\put(100,280){\circle*{12}}
\put(250,280){\circle*{12}}
\put(155,35){\shortstack{\Large (a)}}
\put(375,130){\line(1,0){250}}
\put(375,280){\line(1,0){250}}
\multiput(425,130)(41.666,41.666){4}{\line(1,1){25.001}}
\multiput(425,280)(41.666,-41.666){4}{\line(1,-1){25.001}}
\put(425,130){\circle*{12}}
\put(575,130){\circle*{12}}
\put(425,280){\circle*{12}}
\put(575,280){\circle*{12}}
\put(480,35){\shortstack{\Large (b)}}
\put(700,130){\line(1,0){50}}
\put(700,280){\line(1,0){50}}
\put(900,130){\line(1,0){50}}
\put(900,280){\line(1,0){50}}
\put(750,130){\line(0,1){150}}
\put(900,130){\line(0,1){150}}
\multiput(750,130)(17.5,0){9}{\line(1,0){10}}
\multiput(750,280)(17.5,0){9}{\line(1,0){10}}
\put(750,130){\circle*{12}}
\put(900,130){\circle*{12}}
\put(750,280){\circle*{12}}
\put(900,280){\circle*{12}}
\put(805,35){\shortstack{\Large (c)}}
\end{picture}
\caption{Infrared divergent box graphs in the truncated MSSM.}\label{f4}
\end{figure}

\begin{figure}[htb]
\setlength{\unitlength}{.4pt}
\begin{picture}(972,450)
\thicklines
\put(200,300){\line(-1,1){150}}
\put(200,300){\line(-1,-1){150}}
\put(350,300){\line(1,1){150}}
\put(350,300){\line(1,-1){150}}
\put(200,300){\circle*{12}}\put(350,300){\circle*{12}}
\multiput(200,300)(17.5,0){9}{\line(1,0){10}}
\put(100,200){\circle*{12}}\put(100,400){\circle*{12}}
\multiput(100,200)(0,17.273){12}{\line(0,1){10}}
\put(600,150){\line(1,0){300}}
\put(750,300){\line(-1,1){150}}
\put(750,300){\line(1,1){150}}
\put(750,150){\circle*{12}}
\put(750,300){\circle*{12}}
\put(650,400){\circle*{12}}
\put(850,400){\circle*{12}}
\multiput(650,400)(17.273,0){12}{\line(1,0){10}}
\multiput(750,150)(0,17.5){9}{\line(0,1){10}}
\thinlines
\put(250,35){\shortstack{\Large (a)}}
\put(730,35){\shortstack{\Large (b)}}
\end{picture}
\caption{More infrared divergent one-loop graphs.}\label{f5}
\end{figure}
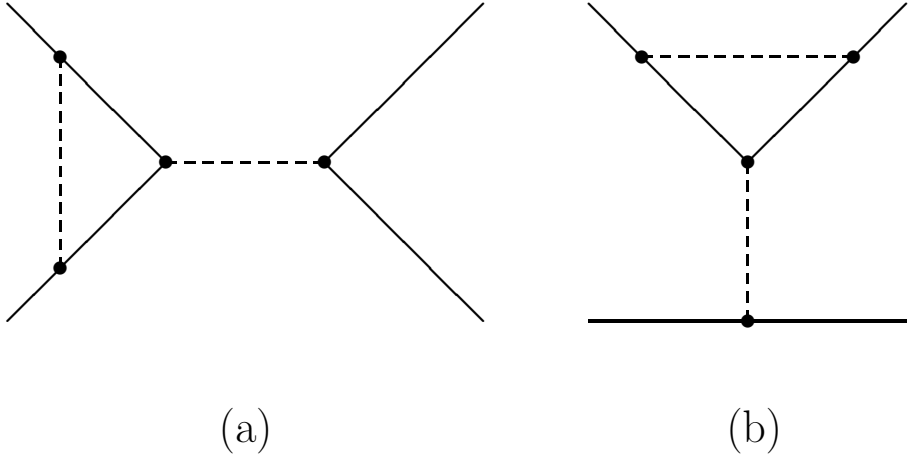

The IR divergent graphs are shown on Figs.~\ref{f4},\ref{f5}. In addition to 
 these, one must include tree graphs and add all $h$ mass insertions to them;
these graphs turn out to be IR finite though. The calculation of each
individual IR divergent part is too complicated to explain here; we briefly
describe one of them (the one corresponding to Fig.~\ref{f4}a in the Appendix).
We simply quote the result in terms of $I_{graph}$:
\begin{equation}
i{\cal M}_{graph}=-\pi^2\cdot g^4\cdot
\log{\frac{\Lambda}{m}}\cdot I_{graph}
\end{equation}

and
\subeqn{
I_{(4a)}&=&\frac{1}{E^2\beta t}\cdot\left(\log{\frac{1+\beta}{1-\beta}}
-i\pi\right)+(t\leftrightarrow u)\label{eI.1}\\
I_{(4b)}&=&-\frac{4}{t}\cdot\phi(u)+(t\leftrightarrow u)\label{eI.2}\\
I_{(4c)}&=&-\frac{4}{s}\cdot\phi(t)+(t\leftrightarrow u)\label{eI.3}\\
I_{(5a)}&=&\frac{1}{E^2\beta s}\cdot\left(\log{\frac{1+\beta}{1-\beta}}
-i\pi \right)\label{eI.4}\\
I_{(5b)}&=&-\frac{2}{t}\cdot\phi(t)+(t\leftrightarrow u)\label{eI.5}
}

with \(E=\frac{1}{2}\sqrt{s}\); $\beta$ is the relative velocity
\(\beta=\sqrt{1-\frac{4M^2}{s}}\) and
\begin{equation}
\phi(x)\equiv\frac{4}{\sqrt{x(x-4M^2)}}\log{
\frac{\sqrt{4M^2-x}+\sqrt{-x}}{\sqrt{4M^2-x}-\sqrt{-x}}}\ .
\end{equation}

The sum of all these terms gives (with spacial momentum $p=\beta\cdot E$)
\begin{equation}
I_{total}=\left(\frac{1}{t}+\frac{1}{u}+\frac{1}{s}\right)\cdot
\left\{\frac{1}{Ep}\cdot\left(\log{\frac{E+p}{E-p}}-i\pi\right)
-\phi(t)-\phi(u)
\right\}\label{eItot}\ .
\end{equation}

This formula eventually coincides with what one gets from
Eqns.~(\ref{e4},\ref{eFandG}). The tree level amplitude is
\begin{equation}
i{\cal M}_{tree}=(2\pi g)^2\cdot\left(\frac{1}{t}+\frac{1}{u}+
\frac{1}{s}\right) \end{equation}
and explicit use of Eqn.~(\ref{eFandG}) allows us to arrive at the same
Eqn.~(\ref{eItot}).

\begin{figure}[htb]
\setlength{\unitlength}{.4pt}
\begin{picture}(972,500)
\thicklines
\put(75,150){\line(1,0){350}}
\put(75,350){\line(1,0){350}}
\put(150,150){\circle*{12}}
\put(350,150){\circle*{12}}
\put(150,350){\circle*{12}}
\put(350,350){\circle*{12}}
\multiput(150,150)(0,17.273){12}{\line(0,1){10}}
\multiput(350,150)(0,17.273){12}{\line(0,1){10}}
\thinlines
\put(240,35){\shortstack{\Large (a)}}
\put(730,35){\shortstack{\Large (b)}}
\put(200,355){\shortstack{\Large M}}
\put(200,115){\shortstack{\Large M}}
\put(25,335){\shortstack{\Large M}}
\put(450,335){\shortstack{\Large M}}
\put(25,135){\shortstack{\Large M}}
\put(450,135){\shortstack{\Large M}}
\put(115,240){\shortstack{\Large m}}
\put(355,240){\shortstack{\Large m}}
\multiput(280,110)(0,56){6}{\rule{3\unitlength}{40\unitlength}}
\put(255,440){\shortstack{\Large cut}}
\thicklines
\put(550,300){\line(1,0){400}}
\put(600,150){\line(0,1){300}}
\put(700,250){\shortstack{${\mbox{\Large 4}}{\mbox{\Large M}}^{\mbox{
\footnotesize 2}}$}}
\put(800,250){\shortstack{${\mbox{\Large 4}}{\mbox{\Large M}}^{\mbox{
\footnotesize 2}}{\mbox{\Large -}}{\mbox{\Large t}}$}}
\put(730,300){\circle*{20}}
\put(850,300){\circle*{20}}
\put(950,300){\oval(600,150)[l]}
\put(790,225){\vector(-1,0){20}}
\put(770,375){\vector(1,0){20}}
\put(730,300){\rule{220\unitlength}{4\unitlength}}
\end{picture}
\caption{Singularities of the graph (\protect\ref{f4}a) in
the complex $s$ plane for fixed $t<0$.}\label{f6}
\end{figure}
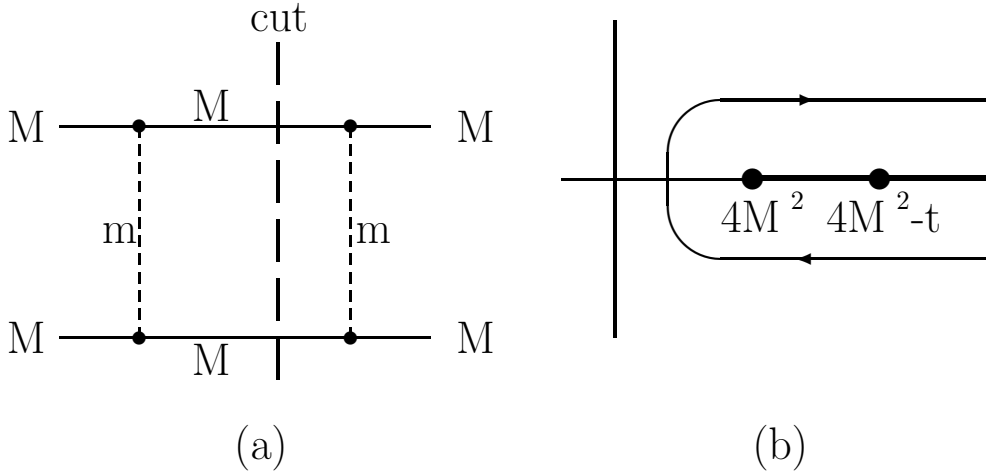

As an alternative way of calculation, we computed $I_{(4a)}$ by a dispersive
calculation. In this calculation we used the usual Landau rules to show that 
this graph has no other singularities at fixed $t<0$ in the complex $s$ plane
but a \mbox{two-$H$} cut from $4M^2$ to $+\infty$, corresponding to the cut
graph on Fig.~\ref{f6}. The singularity for \(4M^2\leq s\leq 4M^2-t\) is there
in spite of the fact that the physical region for this process (we have fixed
$t<0$!) starts at \(s\geq 4M^2-t>4M^2\). The discontinuity across this cut is
calculated from the Cutkosky rules~\cite{Cutk}
\begin{equation}
Im\left\{i{\cal M}_{(1a)}\right\}=g^4\cdot
\int d^4k\cdot\frac{(-i\pi)\delta^+[k^2-M^2]\cdot(-i\pi)\delta^+[(p-k)^2-M^2]}
{\left[(k-p_1)^2-m^2\right]\cdot\left[(k-p_2)^2-m^2\right]}.
\end{equation}

A straightforward calculation leads to
\begin{equation}
Im\left\{i{\cal M}_{(1a)}\right\}=-\pi^2\cdot g^4 \cdot\frac{2\pi} 
{4Ep^3(1-\cos{\Theta})}\cdot \log{\frac{E}{m}}+O(1).\label{eim}
\end{equation}

The real part of the amplitude is determined from the dispersion relation
\begin{equation}
i{\cal M}_{(box)}=\frac{1}{\pi}\cdot\int_{4M^2}^\infty
\frac{ds^\prime}{s^\prime-s-i\varepsilon}\cdot Im\left\{i{\cal M}_{(1a)}
(s^\prime,t) \right\}.\label{edisp}
\end{equation}

We note that no subtraction is needed and that we must include the 
contribution from unphysical \(s^\prime<4M^2-t\) where $p$ is imaginary.
Substituting Eqn.~(\ref{eim}) into Eqn.~(\ref{edisp}) some elementary   
complex integrations lead to Eqn.~(\ref{eI.1}).

This calculation, especially the coincidence of Eqns.~(\ref{e4},\ref{eFandG})
and (\ref{eItot}) shows, through a particular example, the correctness of what
we did in~Sect.~\ref{sWeinb}.

\section{\bf Conclusion}

We have shown, both by general argument and explicit example, that in
diagrams involving scalar loops no powerlike infrared singularities occur.
The introduction of spin does not alter this conclusion, and we use this
result to show that the power coupling proof of the Equivalence Theorem
put forward in~\cite{GrKet} {\it does not suffer from IR divergences}.
However, our reasoning allows us to arrive at this conclusion only in the
massive sector, because the above analysis can break down when massless
propagators are included into loop integrations (for our purposes, gauge
bosons and/or light Higgses of the MSSM model are considered 'massless'.) 

On the basis of these results we put forward the conjecture that, at least
when the light particles are not coupled to each other --- no worse IR
divergences occur than \( O\left(g^n\cdot\log^n{\frac{\Lambda}{m}}\right)\);
we actually prove this  in Sect.~\ref{sWeinb}. A general proof seems
straightforward. The generalization for coupled light particles (which is
certainly the case in the SM) needs a separate investigation. This case is
in a sense similar to the analysis of IR divergences in QCD. One particular
graph that could cause trouble is the same as the one one Fig.~\ref{fig0},
with light particles on the internal lines. Yang-Mills theories certainly
have bad IR behavior which ultimately leads to confinement; the corresponing
investigation might involve an adaptation the methods used there.

\section*{\bf Appendix: The calculation of $I_{(4a)}$}

In this Appendix, as an illustration to the main ideas in all these similar
calculations, we briefly describe the calculation of the IR divergent part in
the box graph of Fig.~\ref{f5}a. The usual Feynman parameter integral is
\begin{equation}
i{\cal M}_{(1a)}=-\pi^2\cdot g^4\cdot
{\int\!\!\int\!\!\int}_0^{x+y+z<1}\frac{dx dy dz}
{\left\{J_m(x,y,z)-i\varepsilon\right\}^2}
\end{equation}
with
\begin{eqnarray}
J_m(x,y,z)&=&m^2\cdot(x+y)+M^2\cdot(1-x-y)^2\\
&&\mbox{}-s\cdot z(1-x-y)-t\cdot xy\ .\nonumber
\end{eqnarray}

\begin{figure}[htb]
\setlength{\unitlength}{.4pt}
\begin{picture}(972,500)
\thicklines
\put(100,400){\line(1,0){350}}
\put(150,150){\line(0,1){300}}
\put(150,400){\circle*{12}}
\put(350,400){\circle*{12}}
\put(150,200){\circle*{12}}
\put(350,200){\circle*{12}}
\multiput(150,402)(52,0){6}{\rule{40\unitlength}{2\unitlength}}
\put(154,400){\vector(0,-1){100}}
\put(154,400){\line(0,-1){200}}
\put(150,200){\vector(1,0){100}}
\put(150,200){\line(1,0){200}}
\put(350,200){\vector(0,1){100}}
\put(350,200){\line(0,1){200}}
\put(365,185){\shortstack{${\mbox{\Large 1}}{\mbox{\Large -}} {\mbox{\Large
i}}$}} \put(160,410){\shortstack{\Large 0}}
\put(360,410){\shortstack{\Large 1}}
\put(100,185){\shortstack{${\mbox{\Large -}}{\mbox{\Large i}}$}}
\put(200,35){\shortstack{\Large (in v)}}
\put(550,400){\line(1,0){350}}
\put(600,150){\line(0,1){300}}
\put(600,400){\circle*{12}}
\put(800,400){\circle*{12}}
\put(800,200){\circle*{12}}
\multiput(550,402)(52,0){7}{\rule{40\unitlength}{2\unitlength}}
\put(600,400){\vector(1,-1){100}}
\put(600,400){\line(1,-1){200}}
\put(800,200){\vector(0,1){100}}
\put(800,200){\line(0,1){200}}
\put(815,185){\shortstack{${\mbox{\Large 1}}{\mbox{\Large -}} {\mbox{\Large
i}}$}} \put(610,410){\shortstack{\Large 0}}
\put(810,410){\shortstack{\Large 1}}
\put(650,35){\shortstack{\Large (in w)}}
\end{picture}
\caption{Avoiding singularities in the Feynman parameters $v$ and $w$ on the 
complex plane. The dashed line represents the possible positions of
singularities.} \label{f7}
\end{figure}
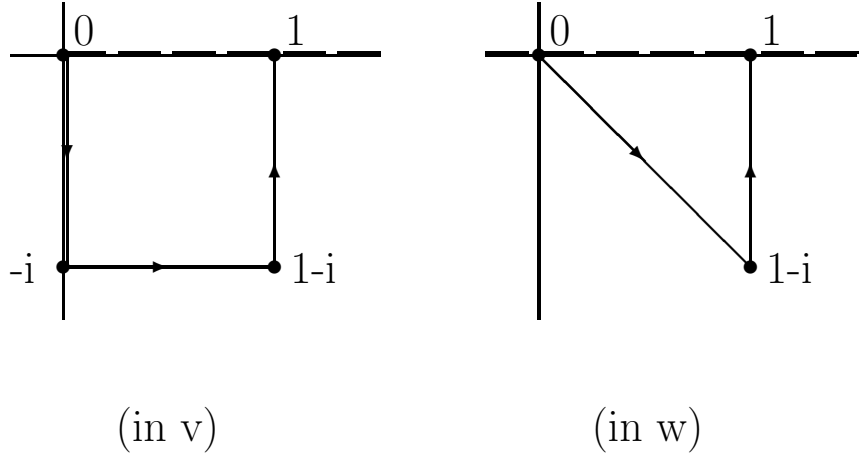

A change of integration variables to $u=1-x-y$, $v=\frac{4 x y}{(x+y)^2}$ and
$w=1-\frac{2z}{1-x-y}$ allows us to have a form
\begin{equation}
J_m=m^2\cdot(1-u)+M^2\cdot u^2+\left(\frac{1-u}{2}\right)^2\cdot
v-s\cdot u^2\cdot\frac{1-w^2}{4}
\end{equation}
in which all the $u$ and $v$ singularities are on the upper complex half 
plane. We  change both contours to ones in the lower half plane as shown on
Fig.~\ref{f7}.

We then treat each of the resulting six integrals in turn. As an example, we 
take the parts with \(v\equiv-i\overline{v}:\ (-i\rightarrow 0)\) and
\(w\equiv 1-i\overline{w}:\ (1-i\rightarrow1)\):
\begin{equation}
i{\cal M}\Longrightarrow -\frac{\pi^2}{2}\cdot g^4\cdot
\int_0^1du\cdot u\cdot(1-u)\cdot\int_0^1
\frac{d\overline{v}}{\sqrt{1-i\overline{v}}}\cdot
\int_0^1\frac{d\overline{w}}{(J_m-i\varepsilon)^2}\end{equation}
with
\begin{eqnarray}
J_m&=&\left[m^2\cdot(1-u)+u^2\cdot(M^2-E^2\cdot\overline{w}^2)\right]\\
&&\mbox{}-i\cdot\left[\left(\frac{1-u}{2}\right)^2\cdot\overline{v}+2E^2\cdot
u^2\cdot\overline{w}\right]\nonumber
\end{eqnarray}
(we set the unit of mass to \(\sqrt{-t}\Rightarrow 1\)).

Next we break up the $u$ integration as \(\int_0^{1/2}du+\int_{1/2}^1du\) and
(using Le\-besgue's theorem) show that the second is IR regular. In the first 
term, a second order Taylor formula for \(\frac{1}{\sqrt{1-i\overline{v}}}\)
allows to drop the remainder term and we are left with (dropping all IR finite
terms) \begin{eqnarray}
i{\cal M}&\Longrightarrow&-\frac{\pi^2}{2}\cdot g^4\cdot
\int_0^{1/2}du\cdot u\cdot(1-u)\cdot\int_0^1d\overline{w}\cdot\\
&&\cdot\int_0^1\frac{d\overline{v}\cdot\left(1+\frac{i}{2}
\overline{v}\right)}
{\left[J_m(\overline{v}=0)-i\varepsilon
-i\cdot\left(\frac{1-u}{2}\right)^2\cdot\overline{v}\right]^2}+
\mbox{IR finite terms}.\nonumber
\end{eqnarray}

An elementary decomposition of the integrand helps us to get rid of the square 
in the denominator
\begin{eqnarray}
i{\cal M}&\Longrightarrow&\frac{\pi^2}{2}\cdot g^4
\cdot\int_0^{1/2}\frac{u\cdot du}{1-u}\cdot\\
&&\ \ \ \ \cdot\int_{0}^{1}d\overline{w}\cdot\int_0^1\frac{d\overline{v}}
{J_m(\overline{v}=0)-i\varepsilon-i\cdot\left(\frac{1-u}{2}\right)^2
\overline{v}}\\ &&+\mbox{IR finite terms.}\nonumber
\end{eqnarray}

We can now do two of the integrals to get
\begin{eqnarray}
i{\cal M}&&\Longrightarrow\ -\frac{i\pi^2}{2E^2}\cdot
 g^4\cdot\int_0^{1/2}\frac{du}{u\cdot(1-u)}\cdot\\
&&\cdot\frac{[\log{(1+i-\sqrt{\ })}-\log{(1+i+\sqrt{\ })}]-
[\log{(i-\sqrt{\ })}-\log{(i+\sqrt{\ })}]}
{\sqrt{\frac{m^2}{E^2}\cdot\frac{1-u}{u^2}-\beta^2-i\varepsilon}}.\nonumber
\end{eqnarray}

Similar calculations for the region \(v:\ (-i\rightarrow 0)\) and
\(w:\ (0\rightarrow 1-i)\) lead to a similar formula with the numerator 
replaced by \[\left[\log{(\sqrt{\ }+1+i)}-\log{(\sqrt{\ }-1-i)}\right],\]
and all other regions end up with no IR divergent contributions. The sum of 
all IR divergent terms, with the variable \(x=\frac{1-u}{u^2}\), is
\begin{eqnarray}
i{\cal M}&\Longrightarrow&-\frac{i\pi^2}{2E^2}\cdot g^4\cdot\int_2^\infty
\frac{dx}{2x}\cdot \left(1+\frac{1}{\sqrt{1+4x}}\right)\cdot\\
&&\cdot\ \frac{\log{(i+\sqrt{\ })}-\log{(i-\sqrt{\ })}+i\pi}
{\sqrt{\mu^2x-\beta^2-i\varepsilon}}\ .\nonumber
\end{eqnarray}

An analysis of the complex $x$ singularities and changing the $x$ contour 
shows that the $\frac{1}{\sqrt{1+4x}}$ term is IR regular. In the other term 
we use \(y=\frac{\mu^2}{\beta^2}\cdot x\)\pagebreak[2] and with similar  
tricks we can show that in the resulting formula,
\begin{eqnarray}
i{\cal M}\Longrightarrow-\frac{i\pi^2}{4Ep}&\cdot&
 g^4\cdot\int_{2\frac{m^2}{p^2}}^\infty\frac{dy}{y}
\cdot\\
&&\cdot\frac{\log{(\frac{i}{\beta}+\sqrt{y-1-i\varepsilon})}
-\log{(\frac{i}{\beta}-\sqrt{y-1-i\varepsilon})}+i\pi}{\sqrt{y-1-i\varepsilon}}
\nonumber
\end{eqnarray}
all the IR divergences are contained in a small neighborhood of zero, in
\pagebreak[1]
an \(\int_{\dots}^\eta dy\) region with {\it any} $m$-independent $\eta>0$.
Because the fraction in the integrand is an
analytic function in \mbox{\(0\leq y\leq\eta\)}, the IR divergence is read
off easily:
\begin{equation}
i{\cal M}\Longrightarrow\frac{\pi^2}{4Ep}\cdot g^4\cdot
\left[i\pi-\log{\frac{1+\beta}{1-\beta}}\right]\cdot\log{\frac{\eta}{\mu^2}},
\end{equation}
which, restoring the unit of \mbox{$\mbox{mass}^2$}, \mbox{$-1\Rightarrow t$},
yields Eqn.~(\ref{eI.1}). 

\section*{\bf Acknowledgements}
The author is indebted to Prof.~J.~F.~Donoghue for the numerous discussions
and helpful comments.

\end{document}